\documentclass[12pt]{article}
\usepackage{graphicx}
\begin{document}
\begin{titlepage}
\title{On the Black Hole Acceleration in the C-metric Space-time}

\author{F. L. Carneiro$\,^{1*}$, S. C. Ulhoa$\,^{2**}$, and
J. W. Maluf$\,^{2***}$, \\
${^1}$ Universidade Federal do Norte do Tocantins, \\
77824-838 Aragua\'ina
TO, Brazil \\
${^2}$ Instituto de F\'{\i}sica, Universidade de Bras\'{\i}lia, \\
70.919-970 Bras\'{\i}lia DF, Brazil}
\date{}
\maketitle
\begin{abstract}
We consider the C-metric as a gravitational field configuration that
describes an accelerating black hole in the presence of a semi-infinite 
cosmic string, along the accelerating direction. We adopt the expression
for the gravitational energy-momentum developed in the teleparallel 
equivalent of general relativity (TEGR) and obtain an explanation 
for the acceleration of the black hole. The gravitational energy enclosed by
surfaces of constant radius around the black hole is evaluated, and in
particular the energy contained within the gravitational horizon is obtained. 
This energy turns out to be proportional to the square root of
the area of the horizon. We find that the gravitational energy of the
semi-infinite cosmic string is negative and dominant for large values of the
radius of integration. This negative energy explains the acceleration 
of the black hole, that moves towards regions of lower gravitational energy 
along the string.
\end{abstract}
\thispagestyle{empty}
\vfill
\noindent (*) fernandolessa45@gmail.com\par
\noindent (**) sc.ulhoa@gmail.com\par
\noindent (***) jwmaluf@gmail.com, wadih@unb.br\par
\end{titlepage}
\newpage

\tableofcontents

\section{Introduction}

The C-metric is a very curious gravitational field configuration. It was
first understood as a solution of Einstein's equations that describes an
accelerating black hole. Nowadays, 
it is clear that the line element describes
not only a pair of black holes accelerated in opposite directions, but a 
sequence of pairs of black holes. A semi-infinite cosmic string is assumed
to be attached to each one of the black holes, along which the black holes 
are accelerated. The acceleration of the black holes is generally supposed to
be due to these semi-infinite cosmic strings, but how exactly these strings
act on the black hole is not clear. 

In this article we present an explanation for this acceleration. The 
explanation is based on energy considerations. We consider the expression for
the gravitational energy-momentum established in the teleparallel equivalent
of general relativity (TEGR), and evaluate the gravitational energy enclosed 
by surfaces of constant radius $R$, such that $R$ lies 
between the gravitational and acceleration horizons. This energy expresses 
both the energy of the black hole and of the cosmic string. 
The gravitational energy density due to the semi-infinite cosmic string only 
is negative (the semi-infinite cosmic string is characterised  by both
the mass parameter and the acceleration parameter), 
and is dominant for large values of the
radius of integration $R$. The total gravitational energy (black
hole plus the cosmic string) is negative and decreases with increasing values
of the radius $R$, as we will show. Assuming that the physical systems in 
nature move towards states of lower energy, the results obtained here 
suggest that the black hole is dragged to regions of lower energy state
along the semi-infinite cosmic string. The absolute value of the (negative)
gravitational energy density 
of the cosmic string increases along the negative $z$ axis. We also
obtain the energy contained within the gravitational horizon. This is the 
energy that cannot escape from the horizon, and is related to the 
irreducible mass of the black hole. It turns out that this energy is 
proportional to the square root of the area of the black hole horizon. 

We will establish a set of tetrad fields adapted to observers that 
accelerate together with the black hole. The dynamical features of this 
frame are investigated by means of the acceleration tensor. This tensor 
will be reviewed in Section 3, and applied to the C-metric black hole in 
Section 4. The outcome of the analysis of the acceleration tensor will help
to understand that the semi-infinite cosmic string is indeed the 
configuration that accelerates the C-metric black hole, as demonstrated in
section 6.

\section{Description of the C-metric}
The C-metric is an exact vacuum 
solution of Einstein's equations that depends on
three parameters: $m$, $\alpha$ and $C$. When $\alpha=0=C$, the 
metric describes the Schwarzschild solution. The parameter $\alpha$ is 
related to acceleration, and $C$ to a deficit angle, or conical singularity.
The exposition below is based on the presentations of 
Refs. \cite{Griffiths,Perlick}. In the latter references, one finds the 
history of this solution, starting with Levi-Civita in 1918
\cite{Levi-Civita}, and ending up with the work of Ehlers and Kundt 
\cite{Ehlers} (see also Ref. \cite{Bicak}). 
The metric is interpreted as describing an
infinite sequence of alternating black holes and asymptotically flat regions
\cite{Griffiths}, and each asymptotically flat region is related to a pair
of causally disconnected black holes. The black holes in each pair are 
supposed to accelerate away from each other along an axis of symmetry of the
space-time. This axis of symmetry contains a conical singularity that may be
physically interpreted as a cosmic string, that is related to the 
acceleration of the black hole. One does not expect this multitude of
pairs of black holes to be realized in 
nature. The very concept of acceleration of a black hole is not of
straightforward comprehension. Nevertheless, this metric will be used here
to model the acceleration of a single astrophysical black hole within a
physical region that can be identified with the surroundings of an 
ideal observer. Such model could eventually describe features
of the outcome of the merger of two black holes that are presently considered
in the generation of the recently observed gravitational waves. However, here
we will be mostly interested in the conceptual issues related to the 
characteristics of an accelerated black hole.

In spherical type coordinates $(t,r,\theta,\Phi$), the line element of the 
C-metric space-time reads

\begin{equation}\label{11}
ds^{2}=\frac{1}{(1+\alpha r \cos{\theta})^{2}}\left(-fdt^{2}
+\frac{1}{f}dr^{2}
+\frac{1}{g}r^{2}d\theta^{2}+gr^{2}\sin^{2}{\theta}d\Phi^{2}\right)\,,
\end{equation}
where $f=f(r)=(1-\alpha^{2}r^{2})(1-2m/r)$,  
$g=g(\theta)=1+2\alpha m\cos{\theta}$, and $(m,\alpha)>0$. 
The physical solutions that we will consider satisfy $0<2\alpha m<1$. It is 
easy to see that when $\alpha=0$, we arrive at the line element of the
Schwarzschild space-time. The angular coordinate $\Phi$ varies in the 
interval $-C\pi<\Phi<C\pi$.

By drawing a small circle around the half-axis $\theta=\pi$, with $(t,r)$ 
constant, we obtain \cite{Griffiths}

\begin{equation}
{{\texttt{circumference}}\over {\texttt{radius}}}=2\pi C(1-2\alpha m)\,,
\label{12}
\end{equation}
which implies the existence of a conical singularity, 
and doing the same around the half-axis $\theta=0$, we find

\begin{equation}
{{\texttt{circumference}}\over {\texttt{radius}}}=2\pi C(1+2\alpha m)\,,
\label{13}
\end{equation}
which also implies the existence of a conical singularity, but with a
different conicity. We choose to eliminate the excess of angular variation
around the upper half-axis $\theta=0$ by fixing the constant $C$ to satisfy
$C=(1+2\alpha m)^{-1}$. In this way, the deficit angles at the half-axis
$\theta=0$ and $\theta=\pi$ are \cite{Griffiths}

\begin{equation}
\delta_{\theta=0}=0\,, \ \ \ \ \ \
\delta_{\theta=\pi} = {{8\pi \alpha m}\over {1+2\alpha m}}\,,
\label{14}
\end{equation}
respectively. The negative $z$ axis is then identified with the 
semi-infinite cosmic string. This semi-infinite cosmic string makes sense 
only if $m \ne 0$ and $\alpha \ne 0$.
Finally, we define the coordinate $\phi$ such that 
$\Phi=C\phi$, where $-\pi<\phi<\pi$, and arrive at the final form of the
line element,

\begin{equation}\label{15}
ds^{2}=\frac{1}{(1+\alpha r \cos{\theta})^{2}}\left(-fdt^{2}
+\frac{1}{f}dr^{2}
+\frac{1}{g}r^{2}d\theta^{2}+
\frac{ g r^{2} \sin^{2}{\theta} }{ (1+2 \alpha m)^{2} } d\phi^{2}
\right)\,.
\end{equation}
The functions $f$ and $g$ are the same as in Eq. (\ref{11}).

The C-metric space-time has a curvature singularity at $r=0$, and two 
coordinate singularities: at $r=2m$, that yields the event horizon $H_{g}$,
and at $r=1/\alpha$, that yields the acceleration horizon $H_a$. 
Thus, ignoring
analytic extensions, the space-time may be divided in three regions 
\cite{Griffiths}: I) $0<r<2m$, which is the interior of the black hole
(non-static region); II) $2m<r<1/\alpha$ (static region); III) 
$1/\alpha<r<\infty$ (non-static region). The coordinates in Eqs.
(\ref{11}) and (\ref{15}) are suitable to Region II, which is the region of
interest to the present analysis. The maximal analytic extension of these
coordinates yields the description of a pair of accelerating black holes in
opposite directions, each of them being in space-time regions that
are causally disconnected. 

The limit $m\rightarrow 0$ of the C-metric is taken by first considering 
Eqs. (\ref{12}) and (\ref{13}), and by noting that the coordinates $\Phi$
in Eq. (\ref{11}) and $\phi$ in Eq. (\ref{15}) are related by $\Phi=C\phi$.
The positive and negative half $z$ axes have now the same angular deficit,
and the line element reduces to 

\begin{eqnarray}\label{16}
ds^{2}&=&\frac{1}{(1+\alpha r \cos{\theta})^{2}}\biggl[-(1-\alpha^{2}r^{2})
dt^{2} \nonumber \\
&+&\frac{1}{1-\alpha^{2}r^{2}}dr^{2}+r^{2}d\theta^{2}+C^2 r^{2}
\sin^{2}{\theta}d\phi^{2}\biggr]\,.
\end{eqnarray}
If the acceleration parameter $\alpha$ vanishes, the line element above can be
transformed into the standard form $ds^2=-dt^2+d\rho^2+ \beta^2\rho^2d\phi^2+dz^2$
of a conical defect in cylindrical coordinates, provided we identify $\beta= C$.

By means of a suitable coordinate transformation, the space-time described
by the line element (\ref{16}) 
can be transformed into the uniformly accelerated
Rindler space-time in cylindrical coordinates. However, this coordinate
transformation cannot be carried out globally, because the topological
defect on the $z$ axis is eliminated by such a global
transformation.

In summary, we see that the C-metric space-time described by Eq. 
(\ref{11}) is a non-linear superposition of a static black hole space-time 
and of a semi-infinite cosmic string along the negative z-axis. 

\section{The acceleration tensor}

In this section we will make a brief presentation of the acceleration
tensor, in order to characterise the acceleration of frames adapted to 
observers in the C-metric space-time. The tetrad field and the inverse frame
field are denoted by $e^a\,_\mu$ and $e_a\,^\mu$, respectively.
[Notation: $a$ and $\mu$ are SO(3,1) and space-time indices, respectively. 
The time and space components are denoted as $a=((0),(i))$ and $\mu=(0,i)$.
The metric tensor $g_{\mu\nu}$ and the flat, tangent space metric tensor
$\eta_{ab}=(-1,+1,+1,+1)$ are related by 
$e^a\,_\mu e^b\,_\nu \eta_{ab}=g_{\mu\nu}$.]

Along an arbitrary
timelike worldline $C$, the velocity of an observer is denoted by $U^\mu$. 
We identify this velocity with the timelike component of the frame field, 
$U^\mu=e_{(0)}\,^\mu$. The acceleration of the observer along this worldline
is defined by the covariant derivative of $U^\mu$ along $C$,

\begin{equation}
a^\mu= {{D U^\mu}\over{d\tau}} ={{De_{(0)}\,^\mu}\over {d\tau}} =
\,U^\alpha \nabla_\alpha e_{(0)}\,^\mu\,, 
\label{17}
\end{equation}
where $\tau$ is the proper time of the observer along $C$, and 
the covariant derivative is constructed out of the Christoffel symbols.
We have considered $U^\alpha=dx^\alpha/d\tau$ along $C$.
Thus, $e_a\,^\mu$ yields the velocity and acceleration of an observer along 
the worldline. Therefore, a given set of tetrad fields, for which 
$e_{(0)}\,^\mu$ describes a congruence of timelike curves, is adapted to a 
particular class of observers, namely, to observers characterized by the 
velocity field $U^\mu=\,e_{(0)}\,^\mu$, endowed with acceleration $a^\mu$. 
If $e^a\,_\mu \rightarrow \delta^a_\mu$ in the limit 
$r \rightarrow \infty$, in an asymptotically flat space-time, then 
$e^a\,_\mu$ is adapted to static observers at spacelike infinity.

An alternative characterization of tetrad fields as an observer's frame may 
be given by considering the acceleration of the whole frame along an 
arbitrary path $x^\mu(\tau)$ of the observer. The acceleration 
of the whole frame is determined by the absolute derivative 
(constructed out of the Levi-Civita connection) of $e_a\,^\mu$
along $x^\mu(\tau)$. Thus, assuming that the observer carries an orthonormal 
tetrad  frame $e_a\,^\mu$, the acceleration of the frame along the path is 
given by \cite{Mashh2,Mashh3,Maluf81,Maluf82}

\begin{equation}
{{D e_a\,^\mu} \over {d\tau}}=\phi_a\,^b\,e_b\,^\mu\,,
\label{18}
\end{equation}
where $\phi_{ab}$ is the antisymmetric acceleration tensor. As discussed in 
Refs. \cite{Mashh2,Mashh3}, in analogy with the Faraday tensor we may identify
$\phi_{ab} \leftrightarrow\ ({\bf a}, {\bf \Omega})$, where ${\bf a}$ is the 
translational acceleration ($\phi_{(0)(i)}=a_{(i)}$) and ${\bf \Omega}$ is the
frequency of rotation of the local spatial frame  with respect to a 
non-rotating, Fermi-Walker transported frame. 
It follows from Eq. (\ref{18}) that

\begin{equation}
\phi_a\,^b= e^b\,_\mu {{D e_a\,^\mu} \over {d\tau}}=
e^b\,_\mu \,U^\lambda\nabla_\lambda e_a\,^\mu\,.
\label{19}
\end{equation}

The acceleration vector $a^\mu$ may be projected on a frame in order to yield

\begin{equation}
a^b= e^b\,_\mu a^\mu=e^b\,_\mu U^\alpha \nabla_\alpha
e_{(0)}\,^\mu=\phi_{(0)}\,^b\,.
\label{20}
\end{equation}
Thus, $a^\mu$ and $\phi_{(0)(i)}$ are not different translational 
accelerations of the frame. The expression of $a^\mu$ given by Eq. (\ref{17})
may be rewritten as

\begin{eqnarray}
a^\mu&=& U^\alpha \nabla_\alpha e_{(0)}\,^\mu 
=U^\alpha \nabla_\alpha U^\mu =
{{dx^\alpha}\over {d\tau}}\biggl(
{{\partial U^\mu}\over{\partial x^\alpha}}
+\,\,^0\Gamma^\mu_{\alpha\beta}U^\beta \biggr) \nonumber \\
&=&{{d^2 x^\mu}\over {d\tau^2}}+\,\,^0\Gamma^\mu_{\alpha\beta}
{{dx^\alpha}\over{d\tau}} {{dx^\beta}\over{d\tau}}\,,
\label{21}
\end{eqnarray}
where $\,\,^0\Gamma^\mu_{\alpha\beta}$ are the Christoffel symbols.
We see that if $U^\mu=\,e_{(0)}\,^\mu$ represents a geodesic
trajectory, then the frame is in free fall and 
$a^\mu=\phi_{(0)(i)}=0$. Therefore we conclude that non-vanishing
values of the latter quantities represent inertial accelerations
of the frame.

An alternative expression of the acceleration tensor is given by
\cite{Maluf81,Maluf82}

\begin{equation}
\phi_{ab}={1\over 2}\lbrack T_{(0)ab}+T_{a(0)b}-T_{b(0)a}\rbrack\,.
\label{22}
\end{equation}
where 

\begin{equation}
T_{abc}=e_b\,^\mu e_c\,^\nu T_{a\mu\nu}=
e_b\,^\mu e_c\,^\nu (\partial_ \mu e_{a\nu}-\partial_\nu e_{a\mu})\,.
\label{23}
\end{equation}
The tensor $\phi_{ab}$ is invariant under coordinate transformations and 
covariant under global SO(3,1) transformations, but not under
local SO(3,1) transformations. Because of this property, $\phi_{ab}$ may be
used to characterise the inertial state of the frame. If the frame is 
maintained static in space-time, then the six components of the tensor 
$\phi_{ab}$ must cancel the six components of the gravitational acceleration
on the frame.

\section{Inertial accelerations in the C-metric space-time}
We will establish a set of tetrad fields constructed in terms of the 
coordinates $(t,r,\theta,\phi)$, whose origin coincide with the centre of the
accelerating black hole. Therefore, the tetrad field below is adapted to 
observers that see the accelerating black hole at rest, and yields the line
element (\ref{15}). It reads

\begin{equation}\label{24}
e_{a\mu}=\Delta\left(
\begin{array}{cccc}
 -A & 0 & 0 & 0 \\
 0 & \frac{\cos \phi  \sin \theta }{A} & \frac{r
   \cos \theta  \cos \phi }{B} & -\frac{r B \sin
   \theta  \sin \phi }{1+2 \alpha  m} \\
 0 & \frac{\sin \theta  \sin \phi }{A} & \frac{r
   \cos \theta  \sin \phi }{B} & \frac{r B \cos
   \phi  \sin \theta }{1+2 \alpha  m} \\
 0 & \frac{\cos \theta }{A} & -\frac{r \sin
   \theta }{B} & 0 \\
\end{array}
\right)\,,
\end{equation}
where
\begin{eqnarray}
A&=&\sqrt{(1-2m/r)(1-\alpha^{2}r^{2})}\,,\label{25}\\
B&=&\sqrt{1+2\alpha m \cos{\theta}}\,\label{26}\,,\\
\Delta &=&\frac{1}{1+\alpha r \cos{\theta}}\,.\label{27}
\end{eqnarray}
Recall that we are assuming $0<2\alpha m<1$.
Since the black hole is accelerated in the negative $z$ direction, the 
observer is likewise accelerated together with the black hole. 
From the point of view of the frame established by Eq. (\ref{24}), the
observer verifies that the black hole is at rest, i.e., the 4-velocity
of the observer is of the type $U^\mu=e_{(0)}\,^\mu=(U^0,0,0,0)$.

In order to calculate the acceleration tensor out of the frame given by
Eq. (\ref{24}), we need the expressions of the torsion tensor
$T_{\lambda\mu\nu}=e^a\,_\lambda T_{a\mu\nu}$. They are given by

\begin{eqnarray}
T_{001}&=&A \Delta \partial_{r}(\Delta A)\,,\nonumber \\
T_{002}&=&A^{2}\Delta \partial_{\theta}\Delta\,,\nonumber \\
T_{112}&=&-\frac{1}{A^{2}}\Delta\partial_{\theta}\Delta\,,\nonumber\\
T_{212}&=&r\frac{\Delta}{AB^{2}}\left[A\partial_{r}(r\Delta)
-B\Delta\right]\,,\nonumber \\
T_{313}&=&r\sin^{2}{\theta}\frac{B\Delta}{A(1+2m\alpha)^{2}}
\left[rAB\partial_{r}\Delta+(AB-1-2m\alpha)\Delta\right]\,,\nonumber \\
T_{323}&=&r\sin^{2}{\theta}\frac{\Delta}{(1+2m\alpha)^{2}}
\biggl\{[B^{2}\cos{\theta}+B\sin{\theta}\partial_{\theta}B\nonumber \\
&-&(1+2m\alpha)\cos{\theta}]\Delta+B^{2}\sin{\theta}\partial_{\theta}\Delta
\biggr\}\,.
\label{28}
\end{eqnarray}

The non-vanishing components of the acceleration tensor are then easily 
calculated, and read

\begin{eqnarray}
\phi_{(0)(1)}&=&F(r;\alpha)\cos{\phi}\sin{\theta}\,,\label{29}\\
\phi_{(0)(2)}&=&F(r;\alpha)\sin{\phi}\sin{\theta}\,,\label{30}\\
\phi_{(0)(3)}&=&F(r;\alpha)\cos{\theta}-\alpha B  \,,\label{31}
\end{eqnarray}
where

\begin{eqnarray}
F(r;\alpha)& =& \alpha \cos{\theta} B\nonumber \\
&+&\frac{m(1 + r^{2}\alpha^{2})-
r^{3} \alpha^{2}-r\alpha\left[r+m\left(r^{2}\alpha^{2}-3\right)\right]
\cos{\theta} }{r^{2}A}\,.\label{32}
\end{eqnarray}
By combining these quantities, we have
\begin{equation}\label{33}
\vec{a} \equiv \phi_{(0)(1)}\hat{x} + \phi_{(0)(2)}\hat{y} + 
\phi_{(0)(3)}\hat{z} = F(r;\alpha)\hat{r}-\alpha B \hat{z}\,.
\end{equation}

The expression above represents the inertial accelerations that are necessary
to impart to the frame (\ref{24}) in order to satisfy the properties that 
Eq. (\ref{24}) must satisfy. For instance: (i) by making 
$\alpha=0$, we obtain 

\begin{equation}\label{34}
\vec{a} = \frac{m}{r^{2}A}\hat{r}\,,
\end{equation}
which is the outward radial acceleration necessary to compensate the 
attractive radial acceleration due to the black hole (the function
$F(r;\alpha)$ generalises the expression above, for non-vanishing
values of $\alpha$); (ii) the term $-\alpha B \hat{z}$ represents the
component of the acceleration on the frame along the negative direction
of the $z$ axis, since the frame is accelerated together with the black 
hole.

In the absence of the black hole, i.e., in the case $m=0$,
the set of tetrad fields obtained from Eq. ({\ref{16}), that represents a 
frame adapted to observers accelerated along the negative $z$ direction, 
reads
\begin{equation}\label{34-A}
e_{a\mu}=\Delta\left(
\begin{array}{cccc}
 -A & 0 & 0 & 0 \\
 0 & \frac{\cos \phi  \sin \theta }{A} & r
   \cos \theta  \cos \phi  & -r \, C \sin
   \theta  \sin \phi  \\
 0 & \frac{\sin \theta  \sin \phi }{A} & r
   \cos \theta  \sin \phi  & r \, C   \cos
   \phi  \sin \theta  \\
 0 & \frac{\cos \theta }{A} & -r \sin
   \theta  & 0 \\
\end{array}
\right)\,.
\end{equation}
It follows from the expression above that 

\begin{equation}
F(0;\alpha) = \alpha\left[ \cos{\theta} -\frac{ \alpha r+\cos{\theta} }
{\sqrt{1-\alpha^{2}r^{2}}}\right]\,.
\label{35}
\end{equation}
Therefore,
\begin{equation}\label{36}
\vec{a} =  \alpha\left[ \cos{\theta} -\frac{ \alpha r+\cos{\theta} }
{\sqrt{1-\alpha^{2}r^{2}}}\right] \hat{r} -\alpha \hat{z}\,.
\end{equation}
At the centre of the coordinate system, $r=0$, we have 
$\vec{a}_{i}= -\alpha \hat{z}$. The constant $C$, that appears in 
Eq. (\ref{34-A}) and that characterises the cosmic
string, does not affect expressions (\ref{35}) and (\ref{36}) above. In fact, 
these expressions can be obtained directly from (\ref{32}) and (\ref{33}) by
making $m=0$ in the latter equations.}

Finally, we mention that for both sets of tetrad fields, Eqs. (\ref{24}) 
and (\ref{34-A}), the frequency of rotation, given by the $\phi_{(i)(j)}$
components of the acceleration tensor, vanish. Both frames are Fermi-Walker
transported.

\section{ A brief review of the TEGR}

The gravitational energy of the C-metric space-time will be investigated 
in the context of the TEGR. This issue is somehow intricate, because we have,
in fact, an accelerated black hole in the presence of a semi-infinite
cosmic string. Up to a certain extent, we will manage to disentangle these 
two gravitational field configurations.

As in previous presentations, we assume that the space-time geometry is 
established by the tetrad fields $e^a\,_\mu$ only. Thus, the only
possible non-trivial definition for the torsion tensor is given by
$T_{a\mu\nu}=\partial_\mu e_{a\nu}-\partial_\nu e_{a\mu}$, as in 
Eq. (\ref{23}). This quantity is trivially related to the
torsion of the Weitzenb\"{o}ck connection 
$\Gamma^\lambda_{\mu\nu}=e^{a\lambda}\partial_\mu e_{a\nu}$. A geometry
defined solely by the tetrad field is more general than the pure
Riemannian geometry, since one can make use of both the curvature and 
torsion tensors, and of the Weitzenb\"ock and Levi-Civita connections.
Of course, the Riemann-Christoffel and Ricci tensors must
exist in order to establish the equivalence between the TEGR and the 
ordinary metric formulation of general relativity.

In the TEGR, it is
possible to rewrite Einstein's equations in terms of $e^a\,_\mu$ and 
$T_{a\mu\nu}$. The Lagrangian density of the theory is defined by
\cite{Maluf1,Maluf2}

\begin{eqnarray}
L&=& -k e({1\over 4}T^{abc}T_{abc}+{1\over 2}T^{abc}T_{bac}-
T^aT_a) -{1\over c}L_M \nonumber \\
&\equiv& -ke\Sigma^{abc}T_{abc} -{1\over c}L_M\,, 
\label{38}
\end{eqnarray}
where $k=c^3/16\pi G$, $T_a=T^b\,_{ba}$, 
$T_{abc}=e_b\,^\mu e_c\,^\nu T_{a\mu\nu}$ and

\begin{equation}
\Sigma^{abc}={1\over 4} (T^{abc}+T^{bac}-T^{cab})
+{1\over 2}( \eta^{ac}T^b-\eta^{ab}T^c)\;.
\label{39}
\end{equation}
$L_M$ stands for the Lagrangian density for the matter fields. 
The Lagrangian density $L$ is invariant under the global SO(3,1)
group. Invariance under the local SO(3,1) group is verified as long as
we take into account the total divergence that arises in the identity 

\begin{equation}
eR(e) \equiv -e\left({1\over 4}T^{abc}T_{abc} + 
{1\over 2}T^{abc}T_{bac} - T^{a}T_{a}\right)
+ 2\partial_{\mu}(eT^{\mu})\,,
\label{40}
\end{equation}
where $R(e)$ is the scalar Riemannian curvature. However, the field 
equations derived from Eq. (\ref{38}) are covariant under local
SO(3,1) transformations, and are equivalent to Einstein's equations. 
They read

\begin{equation}
e_{a\lambda}e_{b\mu}\partial_\nu (e\Sigma^{b\lambda \nu} )-
e (\Sigma^{b\nu}\,_aT_{b\nu\mu}-
{1\over 4}e_{a\mu}T_{bcd}\Sigma^{bcd} )={1\over {4kc}}e\texttt{T}_{a\mu}\,,
\label{41}
\end{equation}
where
$\delta L_M / \delta e^{a\mu}=e\texttt{T}_{a\mu}$. 

Although the definition of the gravitational energy-momentum is established
in  the Hamiltonian framework, it may also be obtained in the framework
of the Lagrangian formulation defined by
(\ref{38}), according to the procedure of Ref. \cite{Maluf2} (we are now 
assuming $c=1=G$). Equation (\ref{41}) may be rewritten as 

\begin{equation}
\partial_\nu(e\Sigma^{a\lambda\nu})={1\over {4k}}
e\, e^a\,_\mu( t^{\lambda \mu} + \texttt{T}^{\lambda \mu})\;,
\label{42}
\end{equation}
where $\texttt{T}^{\lambda\mu}=e_a\,^{\lambda}\texttt{T}^{a\mu}$ and
$t^{\lambda\mu}$ is defined by

\begin{equation}
t^{\lambda \mu}=k(4\Sigma^{bc\lambda}T_{bc}\,^\mu-
g^{\lambda \mu}\Sigma^{bcd}T_{bcd})\,.
\label{43}
\end{equation}
In view of the antisymmetry property 
$\Sigma^{a\mu\nu}=-\Sigma^{a\nu\mu}$, it follows that

\begin{equation}
\partial_\lambda
\left[e\, e^a\,_\mu( t^{\lambda \mu} + \texttt{T}^{\lambda \mu})\right]=0\,.
\label{44}
\end{equation}
The equation above yields the continuity (or balance) equation,

\begin{equation}
{d\over {dt}} \int_V d^3x\,e\,e^a\,_\mu (t^{0\mu} +\texttt{T}^{0\mu})
=-\oint_S dS_j\,
\left[e\,e^a\,_\mu (t^{j\mu} +\texttt{T}^{j\mu})\right]\,,
\label{45}
\end{equation}
where $S$ is the boundary of an arbitrary 3-dimensional volume $V$.
Therefore we identify
$t^{\lambda\mu}$ as the gravitational energy-momentum tensor \cite{Maluf2},

\begin{equation}
P^a=\int_V d^3x\,e\,e^a\,_\mu (t^{0\mu} 
+\texttt{T}^{0\mu})\,,
\label{46}
\end{equation}
as the total energy-momentum contained within the volume $V$,

\begin{equation}
\Phi^a_g=\oint_S dS_j\,
\, (e\,e^a\,_\mu t^{j\mu})\,,
\label{47}
\end{equation}
as the gravitational energy-momentum flux \cite{Maluf2,Maluf3}, and

\begin{equation}
\Phi^a_m=\oint_S dS_j\,
\,( e\,e^a\,_\mu \texttt{T}^{j\mu})\,,
\label{48}
\end{equation}
as the energy-momentum flux of matter \cite{Maluf3,Maluf3-3}. In view of 
(\ref{42}), Eq. (\ref{46}) may be written as 
$P^a=-\int_V d^3x \partial_j \Pi^{aj}\,,$
from what follows

\begin{equation}
P^a=-\oint_S dS_j\,\Pi^{aj}\,,
\label{49}
\end{equation}
where $\Pi^{aj}=-4ke\,\Sigma^{a0j}$. A summary of all issues discussed above
may found in Ref. \cite{Maluf9}.  

The passage from a volume integral to a
surface integral such as Eq. (\ref{49}) cannot be carried out in the presence 
of singularities (admitting that the space-time has singularities),
and for this reason we consider Eq. (\ref{49}) as the
definition of the gravitational energy-momentum. It must be noted, however, 
that the same feature takes place in the definition of the ADM gravitational
energy-momentum, where the integrals of total divergences are transformed 
into surface integrals. The surface integral 
is superior with respect to the volume integral, because the gravitational 
field on the surface of integration $S$ 
carries information about the interior region, and the integral can be
carried out more easily. In addition, definition (\ref{49}) represents the 
total energy of the space-time within the surface $S$.

Equation (\ref{49}) is the 
definition for the gravitational energy-momentum presented in Ref.
\cite{Maluf4}, obtained in the framework of the vacuum field
equations in Hamiltonian form. It is invariant under coordinate 
transformations of the three-dimensional space and under time 
reparametrizations. Note that (\ref{44}) is a true energy-momentum 
conservation equation. 
In the ordinary formulation of arbitrary field theories, energy, 
momentum, angular momentum and the centre of mass
moment are frame dependent field quantities, that 
transform under the global SO(3,1) group. In particular, the energy 
transforms as the zero component of the energy-momentum four-vector. 
These features of special relativity must also hold in general relativity,
since the latter yields the former in the limit of weak (or vanishing) 
gravitational fields. 

The problem of defining the gravitational energy-momentum has a long
history, and is probably as old as general relativity itself. It is well
known the existence of several expressions of pseudo-tensors, including one
by proposed by Einstein, and all these expressions have an obvious limitation
since they are not tensors. Nowadays, the majority of the objections 
(if not all) against the existence of a 
localized expression for the gravitational energy-momentum is 
justified by invoking the principle of equivalence. The idea is that the
affine connection in general relativity can be made to vanish at a point
in space-time, or even along an arbitrary worldline (timelike or spacelike).
However, as argued before \cite{Babak},  
the vanishing of the affine connection is a feature of differential
geometry, and not a principle of nature. The problem regarding the definition
of the gravitational energy-momentum has to do with transformation of frames,
not transformation of coordinates. 

This whole issue has been thoroughly discussed
in section 5 of Ref. \cite{Maluf9}. In subsection 5.3 of the latter reference,
we have shown that the definitions of energy-momentum and 4-angular momentum
that arise in the TEGR satisfy the Poincar\'e algebra in the phase space
of the theory. This result, together with the calculation of the gravitational
energy contained within the external event horizon of a Kerr black hole
\cite{Maluf4},
distinguishes our definition from all other existing definitions. However,
the gravitational energy-momentum and 4-angular momentum must be frame 
dependent, as we argued above. The tetrad frame may be feely chosen, since
every observer in space-time, along arbitrary timelike worldlines, carries 
his/her own tetrad frame. 

The local SO(3,1) symmetry is not present in 
expressions (\ref{46}) and (\ref{49}) for the gravitational energy-momentum
but, in practice, the latter can be evaluated in any frame in space-time, 
static (with respect to the spacelike infinity), stationary, free-fall, etc.

\section{Gravitational energy in the C-metric space-time}

Expression (\ref{49}) for the total gravitational energy takes into
account altogether the contributions of the black hole and of the 
infinite cosmic string. Both gravitational field configurations are
formally given by the integrand in Eq. (\ref{46}). As we already 
mentioned, the analytical expression of the semi-infinite cosmic string is
not yet known. Therefore, expression (\ref{49}) is better suited to the 
analysis of the total gravitational energy of the C-metric space-time, 
because it incorporates the features of the non-linear superposition of the
two geometrical field configurations.

We will evaluate the gravitational energy of the C-metric space-time
in a region not close to 
the acceleration horizon $H_a$ determined by $r=1/\alpha$. We are interested
in situations of present astrophysical interest, and thus we will ignore
the acceleration horizon and possible maximal extensions of the 
C-metric space-time. In order to evaluate the surface integral in 
Eq. (\ref{49}), we need the 
quantities $\Pi^{aj}=-4k\,\Sigma^{a 0j}$. We find

\begin{eqnarray}
\Sigma^{(0)01}&=&\frac{(1+2m\alpha-2AB +B^{2})\Delta-2rAB\partial_{r}\Delta}
{2rB\Delta^{4}}\,,\label{50}\\
\Sigma^{(0)02}&=&\frac{\Delta[(1+2m\alpha-B^{2})\cot{\theta}
-B\partial_{\theta}B]-2B^{2}\partial_{\theta}\Delta}
{2r^{2}A\Delta^{4}}\,.\label{51}
\end{eqnarray}

The gravitational energy $P^{(0)}$ contained within a surface of constant 
radius $r$ is determined by 

\begin{equation}
P^{(0)}=4k\oint_{S}dS_{1}e\Sigma^{(0)01}\,,
\label{52}
\end{equation}
where $dS_1=d\theta d\phi$. The surface of constant radius $R$ is depicted
in Figure 3 of Ref. \cite{Griffiths}, but as noted in this reference, 
there is no sharp vertex in the negative $z$ axis, at $\theta=\pi$, in spite
of the presence of the semi-infinite cosmic string. The surface is regular at
this point, so that Eq. (\ref{58}) below (where $r=2m$) may be easily 
obtained. By carrying out the integral above on the 
surface of constant radius $r=R$, we obtain

\begin{eqnarray}
P^{(0)}&=&
\frac{1}{2\alpha R^{1/2}(1+2m\alpha)}
\biggl\{-\frac{1}{1-R^{2}\alpha^{2}}
\biggl[\sqrt{R-2mR\alpha}-\sqrt{R+2mR\alpha}  \nonumber\\
&{}&+2\alpha R^{3/2} A-\alpha^{2}R^{2}\biggl(\sqrt{R-2mR\alpha}
-\sqrt{R+2mR\alpha}\biggr)\biggr] \nonumber\\
&{}&-\frac{2[R-m(1-R\alpha)]}{\sqrt{2m-R}}
\biggl(\tan^{-1}{\sqrt{\frac{R+2mR\alpha}{2m-R}}} 
-\tan^{-1}{\sqrt{\frac{R-2mR\alpha}{2m-R}}}\biggr)\biggr\}\,. \nonumber \\
&{}&
\label{53}
\end{eqnarray}
In view of the relation 
$\tan^{-1}z=-\frac{i}{2}\ln{\big(\frac{i-z}{i+z}\big)}$, we have

\begin{eqnarray}
&{}&\frac{1}{\sqrt{2m-R}}\left(\tan^{-1}{\sqrt{\frac{R+2mR\alpha}{2m-R}}}
-\tan^{-1}{\sqrt{\frac{R-2mR\alpha}{2m-R}}}\right)\nonumber\\
&{}&=-\frac{1}{i\sqrt{R-2m}}{\frac{i}{2}}\ln{\bigg[\frac{(\sqrt{2m-R}
+i\sqrt{R-2mR\alpha})(\sqrt{2m-R}-i\sqrt{R+2mR\alpha})}
{(\sqrt{2m-R}-i\sqrt{R-2mR\alpha})(\sqrt{2m-R}+
i\sqrt{R+2mR\alpha})} \bigg]}\nonumber\\
&{}&=-\frac{1}{2\sqrt{R-2m}}\ln{\bigg[\frac{(\sqrt{R-2m}+\sqrt{R-2mR\alpha})
(\sqrt{R-2m}-\sqrt{R+2mR\alpha})}{(\sqrt{R-2m}-\sqrt{R-2mR\alpha})
(\sqrt{R-2m}+\sqrt{R+2mR\alpha})} \bigg]}\,. \nonumber \\
&{}& \label{54}
\end{eqnarray}
After a number of simplifications, we finally arrive at

\begin{eqnarray}
P^{(0)}&=&\frac{1}{2\alpha R^{1/2}(1+2m\alpha)}
\biggl\{-\frac{1}{1-\alpha^{2}R^{2}}\biggl[(\sqrt{R-2m\alpha R}
-\sqrt{R+2m\alpha R}) \nonumber\\
&+&2\alpha R^{3/2} A-
\alpha^{2}R^{2}(\sqrt{R-2m\alpha R}-
\sqrt{R+2m\alpha R})\biggr]\nonumber\\
&+&\frac{[R-m(1-R\alpha)]}{\sqrt{R-2m}}\times \nonumber \\
&\times&\ln \biggl[\frac{(\sqrt{R-2m}+\sqrt{R-2m\alpha R})
(\sqrt{R-2m}-\sqrt{R+2m\alpha R})}{(\sqrt{R-2m}-
\sqrt{R-2m\alpha R})(\sqrt{R-2m}+\sqrt{R+2m\alpha R})} \biggr] \biggr\}\,.
\nonumber \\
&{}&
\label{55}
\end{eqnarray}

The gravitational energy $P^{(0)}_h$ contained within the event horizon
$H_g$ can be evaluated by taking the limit $r\rightarrow 2m$ in the 
expression above. We find

\begin{equation}\label{56}
P^{(0)}_h  =4k\oint dS_{1}
\displaystyle{\lim_{r\to 2m}}(e\Sigma^{(0)01})=
\frac{2m}{(1+2m\alpha)\sqrt{1-2m\alpha}}\,.
\end{equation}
When $\alpha=0$, Eq. (\ref{55}) simplifies to 

\begin{equation}\label{57}
P_{Schw}^{(0)}=R\left(1-\sqrt{1-\frac{2m}{R}}\right)\,,
\end{equation}
which is a well known result (obtained previously in the TEGR and by means
of quasilocal expressions for the gravitational energy) that yields 
$P^{(0)}=m$ in the limit $R\rightarrow \infty$.

It is very interesting to note that the energy contained within the event
horizon $H_g$ given by Eq. (\ref{56}) is related to the area $A_h$ of the 
event horizon calculated in Ref. \cite{Griffiths}. In the latter reference,
the area $A_h$ is shown to be

\begin{equation}
A_h={{16\pi C m^2}\over {1-4\alpha^2m^2}}\,.
\label{58}
\end{equation}
Considering the value of $C$ adopted in Subsection 3.1 (as well as in Ref.
\cite{Griffiths}), $C=(1+2\alpha m)^{-1}$, it is straightforward to obtain 
the relation

\begin{equation}
P^{(0)}_h={\sqrt{A_h}\over {2\pi^{1/2}}}\,.
\label{59}
\end{equation}
This relation may be useful in the study of the thermodynamics of the 
C-metric black hole.

In the limit $m\rightarrow 0$, $C$ is no longer given by 
$C=(1+2\alpha m)^{-1}$, but it can acquire arbitrary values (see 
Eq. (\ref{16}))
\footnote{We remark that when $C=1$, the line element (\ref{16}), and
consequently the set of tetrad fields (\ref{34-A}), does not represent the 
ordinary Minkowski space-time, but just a partition of the latter, delimited
by the acceleration horizons. In section 5 of ref. [1], it is very clearly 
stated that test particles at the origin r = 0 acquire acceleration along
the $\pm z$ directions, which certainly is not a feature of the ordinary,
full Minkowski space-time.}.
The expression 
of $P^{(0)}$ in this limit is obtained directly from the tetrad fields 
(\ref{34-A}), and represents the energy of an infinite cosmic string only,
evaluated in an accelerated frame along the negative $z$ axis. 
It is given by 

\begin{equation}
P^{(0)}_{cs}=-{R C \over {\sqrt{1-\alpha^2 R^2}}} +
\frac{1+C}{4\alpha}\ln{\bigg(\frac{1+ \alpha R}{1-\alpha R}\bigg)}\,.
\label{60}
\end{equation}

\begin{figure}[htbp]
	\centering
		\includegraphics[width=0.80\textwidth]{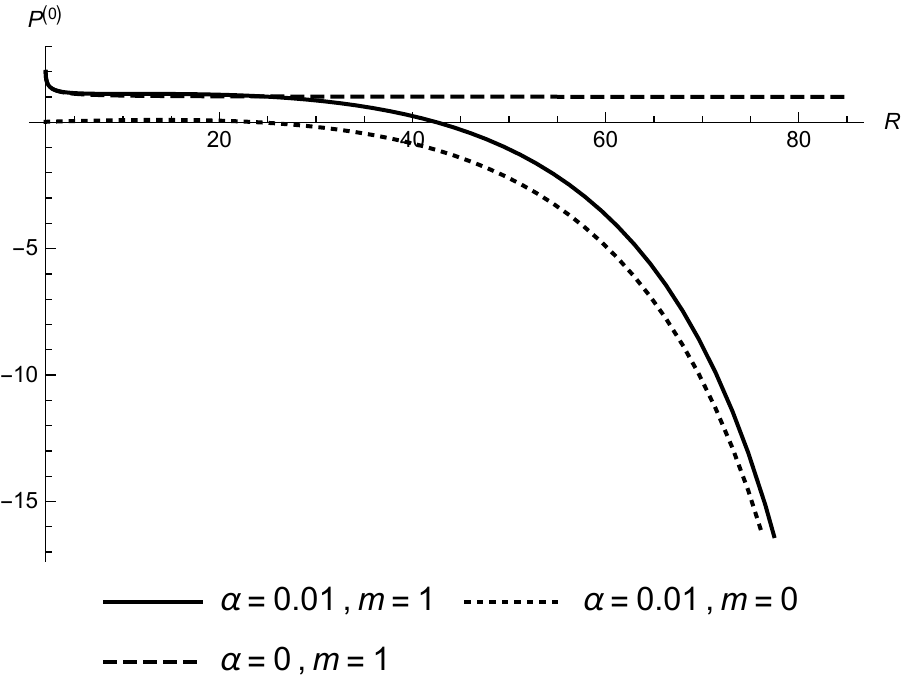}
	\caption{Gravitational energy $P^{(0)}$ given by Eq. (\ref{55}),
	 for various values of $R$ of the surface of integration, considering
	 $R\geq 2m$, $\alpha=0.01$ and $m=1$ in natural units (continuous thick
	 line). The curve for which
	 $\alpha=0$ represents the Schwarzschild black hole (Eq. (\ref{57})), 
	 and the one with $m=0$ represents the infinite cosmic string only
	 (Eq. (\ref{60})). }
	\label{fig5-1}
\end{figure}

\begin{figure}[htbp]
	\centering
		\includegraphics[width=0.80\textwidth]{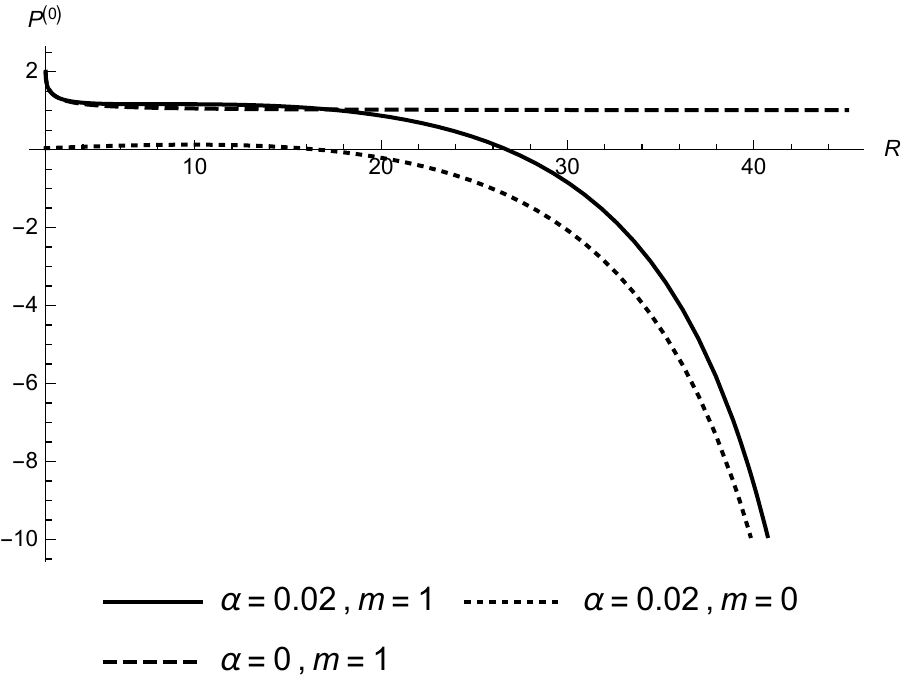}
	\caption{Gravitational energy $P^{(0)}$ given by Eq. (\ref{55}),
	 for various values of $R$ of the surface of integration, considering
	 $R\geq 2m$, $\alpha=0.02$ and $m=1$ in natural units (continuous thick
	 line). The curve for which
	 $\alpha=0$ represents the Schwarzschild black hole (Eq. (\ref{57})), 
	 and the one with $m=0$ represents the infinite cosmic string only
	 (Eq. (\ref{60})). }
	 	\label{fig5-2}
\end{figure}

\begin{figure}[htbp]
	\centering
		\includegraphics[width=0.80\textwidth]{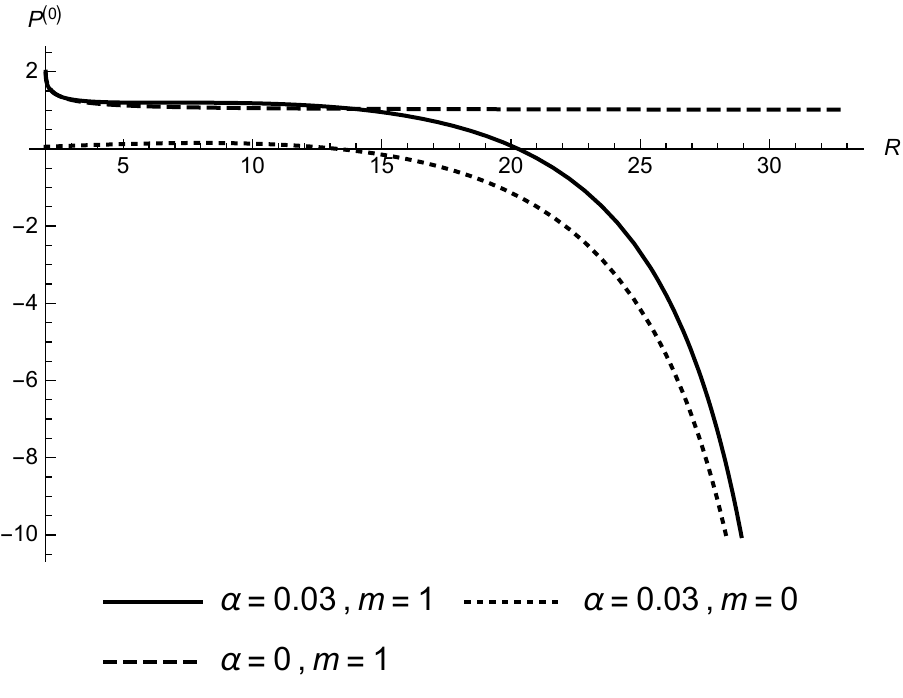}
	\caption{Gravitational energy $P^{(0)}$ given by Eq. (\ref{55}),
	 for various values of $R$ of the surface of integration, considering
	 $R\geq 2m$, $\alpha=0.03$ and $m=1$ in natural units (continuous thick
	 line). The curve for which
	 $\alpha=0$ represents the Schwarzschild black hole (Eq. (\ref{57})), 
	 and the one with $m=0$ represents the infinite cosmic string only
	 (Eq. (\ref{60})). }
	 	\label{fig5-3}
\end{figure}

In the Figures below, we consider expression (\ref{55}) for $P^{(0)}$ 
and display the total gravitational energy enclosed 
by a surface of constant radius $R$, considering $m=1$ in
natural units. In Figure 
\ref{fig5-1} we display altogether: (i) $m=1$, $\alpha=0.01$;
(ii) $m=1$, $\alpha=0$ (Schwarzschild); (iii) $m=0$, $\alpha=0.01$.
In the latter case (iii), we have considered Eq. (\ref{60}) and 
have chosen 
$C=\lbrack 1+2(0.01)\rbrack^{-1}$ in
order to make a consistent comparison with the first two cases, i.e., the 
value of $C$ is the same in the three cases.
In Figures \ref{fig5-2} and \ref{fig5-3}
we consider $\alpha=0.02$ and $\alpha=0.03$, respectively, and the 
corresponding values of $C$. In all cases,
we see that for higher values of the radial coordinate $R$, the energy of 
the infinite cosmic string dominates. 

In Figures \ref{fig5-1}, \ref{fig5-2} and \ref{fig5-3} we see that the energy
in the space-time of a pure infinite cosmic string is negative. As we
mentioned above, this energy dominates when we consider higher volumes of 
integration. Thus, the energy density in regions for higher values of the 
radius of surface integration $R$ (i.e., $R$ approaching $1/\alpha$)
is negative. It is likely that this
negative energy density is responsible for the acceleration 
of the black hole, since the black hole is moving towards the region of 
negative energy density. One argument in support of this conclusion is the 
following. Let us consider the gravitational energy contained within a 
surface of constant radius $r$ in a Schwarzschild space-time. It is given by
Eq. (\ref{57}). By making $r=2m$ and $r \rightarrow \infty$ 
in the latter equation, we obtain
$P^{(0)}=2m$ and $P^{(0)}=m$, respectively. The case $r=2m$ is in agreement
with Eq. (\ref{56}). These results can also be obtained by means of the 
the quasi-local expression for the gravitational energy given by Brown and 
York \cite{Brown}. Thus, in the region between $r=2m$ and 
$r\rightarrow \infty$, the gravitational energy density is negative. The
negative gravitational energy outside the event horizon may be 
identified with the negative Newtonian binding energy, which is attractive,
as noted by Brown and York (see Eq. (6.16) of Ref. \cite{Brown}).
The same feature may be occurring here: a region of 
negative gravitational energy density exerts gravitational attraction, but 
in this case the black hole is being attracted, or accelerated along the
semi-infinite cosmic string in the negative $z$ axis. One may think that the
black hole is approaching a state of lower energy, as do ordinary bodies in 
classical physics.

The energy of space-time (topological) defects may be positive or 
negative, according to an ``addition" or ``removal" of a continuum medium to
the space-time. Cosmic strings are disclination-type defects, and are highly
energetic defects compared to dislocations. This issue is discussed in Refs.
\cite{Katanaev1,Katanaev2}. (See also Eq. (32) of Ref. \cite{Maluf10}, which
presents the energy per unit length of a cosmic string. For a parameter
$\beta_0>1$, this energy is negative.)
When a substantial fraction of a space-time is
``removed", as in the case of the infinite cosmic string (according to
Eq. (\ref{14})), the total
energy of the space-time may be negative in the frame accelerated along
the negative $z$ axis.

In view of Eqs. (\ref{55}) and (\ref{60}), we may identify the energy of
the black hole only as 

\begin{equation}
P^{(0)}_{bh}=P^{(0)}-P^{(0)}_{cs}\,.
\label{61}
\end{equation}
In the evaluation of $P^{(0)}_{cs}$, the constant $C$ is numerically 
chosen to be $C=(1+2\alpha m)^{-1}$, where $m$ and $\alpha$ are the values 
that yield $P^{(0)}$. Thus, both $P^{(0)}$ and $P^{(0)}_{cs}$ 
are endowed with the same constant $C$.

\begin{figure}[htbp]
	\centering
		\includegraphics[width=0.80\textwidth]{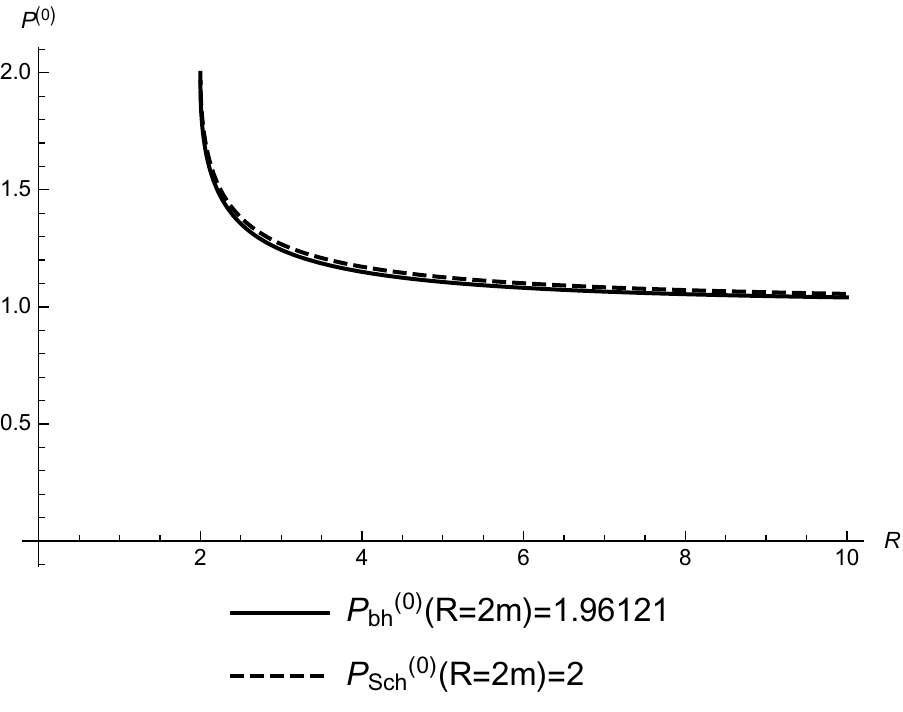}
	\caption{Gravitational energy $P^{(0)}_{bh}=P^{(0)}-P^{(0)}_{cs}$
	     (continuous line),
		 for various values of the surface of integration $R$, 
         considering $m=1$ and $\alpha=0.01$ in natural units. The 
         dashed line represents the Schwarzschild black hole 
         ($\alpha=0$).}
	\label{fig-6-6}
\end{figure}

The identification above
turns out to be consistent, as we see in Figure \ref{fig-6-6}, because the 
difference between this energy and the energy obtained from Eq. (\ref{57}),
that represents the energy enclosed by surfaces of constant radius 
$R$ in the Schwarzschild space-time, is not too much significant. 
Although the surfaces of constant radius $R$ 
are not strictly the same in the space-times with 
and without the acceleration parameter $\alpha$ (i.e., there is no covariant
relation between the radius $R$ in the two situations), the result displayed
by Figure \ref{fig-6-6} is qualitatively relevant to indicate the 
consistency of our analysis.

As a final remark, we note that $P^{(0)}_{cs}$ given by Eq. (\ref{60})
vanishes in the flat space-time limit, which is obtained by requiring 
simultaneously $\alpha =0$ and $C=1$. In addition, the following limits
are verified: (i) when $\alpha \rightarrow 0$, Eq. (\ref{55}) reduces to
the energy of the Schwarzschild space-time, as given by Eq. (\ref{57}); 
(ii) when both $m$ and $\alpha$ vanish, the gravitational energy vanishes;
(iii) when only the mass parameter $m$ vanishes, Eq. (\ref{55}) reduces
to Eq. (\ref{60}), with $C=1$. With respect to the latter equation, we
note that the gravitational energy of the ordinary
Minkowski space-time vanishes, either in the ordinary Cartesian or Rindler
coordinates. However, the space-time represented by Eq. (\ref{16}) is a
partition of the full Minkowski space-time, delimited by the acceleration
horizons. Obviously, such a partition does not
represent the complete, ordinary Minkowski space-time, in the same way
that a partition of a sphere does not represent a sphere.

\section{Conclusions}

In this article we have addressed the C-metric space-time and have presented
an explanation for the acceleration of the black hole. We recall that
the C-metric space-time is a gravitational field configuration that describes
an accelerated black hole along a semi-infinite cosmic string. The black
hole is characterised by the mass parameter $m$, and the acceleration 
$\alpha$ yields the angular deficit $C$ in the negative part of the 
$z$ axis ($\theta = \pi$), characterised by $C=(1+2\alpha m)^{-1}$.

We obtained the expression for the gravitational energy 
contained within a surface of constant radius $R$, around the centre of the
accelerated black hole. In the limit 
$r\rightarrow 2m$, we found the energy contained within the gravitational
horizon, given by Eq. (\ref{56}). This is the energy that cannot escape from
the black hole. This energy may be identified with $2M_{irr}$, where 
$M_{irr}$ is sometimes defined as the irreducible mass of the black hole,
in analogy with the definition of the irreducible mass of the Kerr 
black hole. For large values of the radius of integration $R$, the total 
gravitational energy (black hole plus the infinite cosmic string) 
is negative, according to Figures \ref{fig5-1}, \ref{fig5-2} and 
\ref{fig5-3}. It is clear that this negative energy is dominated by the 
energy of the infinite cosmic string. As we argued at the end of 
Section 6, we may interpret the black hole as being dragged
(accelerated) towards a state of lower energy, along the 
infinite cosmic string. The larger is the value of the radius of
integration $R$, the more negative is the gravitational energy density
of the cosmic string. Therefore, the black hole moves towards regions of
lower gravitational energy density. 

The accelerated black hole, as described by the 
C-metric space-time, is not physically equivalent to the situation where
the black hole is a rest, and the observer undergoes an acceleration
$-\alpha$. In particular, by means of a local Lorentz transformation, we
cannot remove the acceleration of the black hole in the C-metric 
space-time. 

We mention finally that we carried out a local Lorentz transformation on the
set of tetrad fields (\ref{24}) such that the new frame is accelerated in the
positive $z$ direction with acceleration $+\alpha$. This new frame
represents a static (or nearly static) frame in space-time, where the
observer is no longer attached to the black hole. We calculated the 
gravitational energy in this new frame and found that the resulting relation
between $P^{(0)}$ and $R$ is extremely similar to Figures 1, 2 and 3, i.e.,
there is not a single qualitative difference between the relation of $P^{(0)}$
and $R$ in the two situations. This result ensures the frame independence of
our main conclusion, in spite of the quantitative differences  
for the gravitational energy arising in the consideration 
of the nearly static frame, as compared to Eq. (\ref{55}) (i.e., the 
latter equation is not frame independent).
The quantitative differences are due to the emergence of the quantity
$\gamma(t)$ in some terms in the expression of the gravitational energy in the 
nearly static frame, where $\gamma(t)=(1-v(t)^2/c^2)^{-1/2}$.

\end{document}